\documentclass[submitting, floatfix]{nst}
\usepackage{subfigure,dcolumn}
\usepackage[T2A,T1]{fontenc}
\usepackage[english]{babel}
\usepackage{color}
\usepackage{listings}
\usepackage{natbib}
\let\footnote\relax

\let\textcite\relax

\let\citeauthor\relax
\let\citeyear\relax
\expandafter\let\csname
ver@natbib.sty\endcsname\relax

\let\linenumbers\relax

\usepackage[
    backend=biber,
    bibencoding=utf8,
    style=numeric-comp,
    giveninits=true,
    autolang=other, 
    sorting=none,
    doi=true, isbn=false, eprint=false, url=false, date=year, 
    maxcitenames=3,
    mincitenames=3,
    maxbibnames=3,
    minbibnames=3,
]{biblatex}

\usepackage{csquotes}
\renewbibmacro{in:}{}
\newcommand{\figwb}{0.45}

\begin{document}

\renewcommand{\bibliography}[1]{}

\title{Data-driven simultaneous vertex and energy reconstruction for large liquid scintillator detectors}

\author{Gui-hong Huang}
\affiliation{School of Applied Physics and Materials, Wuyi University, Jiangmen 529020, China}
\author{Wei Jiang}
\affiliation{Institute of High Energy Physics, Chinese Academy of Sciences, Beijing 100049, China}
\affiliation{School of Physical Sciences, University of Chinese Academy of Science, Beijing 100049, China}
\author{Liang-jian Wen}
\affiliation{Institute of High Energy Physics, Chinese Academy of Sciences, Beijing 100049, China} 
\author{Yi-fang Wang}
\affiliation{Institute of High Energy Physics, Chinese Academy of Sciences, Beijing 100049, China}
\author{Wu-Ming Luo}
\email[Corresponding author, ]{luowm@ihep.ac.cn}
\affiliation{Institute of High Energy Physics, Chinese Academy of Sciences, Beijing 100049, China}

\begin{abstract}
 High precision vertex and energy reconstruction is crucial for large liquid scintillator detectors such as JUNO, especially for the determination of the neutrino mass ordering by analyzing the energy spectrum of reactor neutrinos. This paper presents a data-driven method to obtain more realistic and more accurate expected PMT response of positron events in JUNO, and develops a simultaneous vertex and energy reconstruction method that combines the charge and time information of PMTs. For the JUNO detector, the impact of vertex inaccuracy on the energy resolution is about 0.6\%.
\end{abstract}


\maketitle

\section{Introduction}
Neutrino studies have brought ground breakthroughs to particle physics and astrophysics. Neutrino experiments Super-Kamiokande and SNO proved that neutrinos have mass~\cite{Super-Kamiokande, SNO}, Kamland, SNO and Daya Bay measured 3 of the oscillation parameters ($\Delta m_{21}^2, \textrm{sin}^2\theta_{12}, \textrm{sin}^2\theta_{13}$) to a precision of percent level~\cite{kamland, SNO2011, dyb}. The detection of high energy cosmic neutrinos by IceCube experiment has opened a new window of neutrino astronomy~\cite{icecube2013}. 
Next-generation detectors aiming to determine the neutrino mass ordering (NMO) and the CP violation phase are under construction.
Jiangmen Underground Neutrino Observatory (JUNO) will be the largest liquid scintillator (LS) detector in the world with the primary goal of determining the NMO~\cite{juno}.
It demands precise reconstruction of reactor neutrinos to extract the NMO information from the energy spectrum.

Reactor neutrinos are detected via inverse beta decay (IBD) in LS, the daughter particles positron and neutron will produce correlated prompt and delayed signals, respectively. The energy of the incident neutrino can be deduced from the energy of the positron.
In LS detectors, the reconstruction of the positron vertex and energy are strongly correlated. On the one hand, due to the non-uniform detector response, the vertex precision will affect the energy non-uniformity, which is one of the main contributing factors to the energy resolution. 
On the other hand, the vertex resolution is highly energy dependent, positrons with larger energy will emit more photons, resulting more accurate reconstruction of the vertex. 
There were quite a few studies on the vertex or energy reconstruction of positrons in JUNO previously, which includes likelihood methods~\cite{Liu_2018,Wu_2019, Li_2021, Huang_2021} and machine learning methods ~\cite{mlVertex1, mlVertex2, mlEnergy1}. The basic strategy of likelihood method is to obtain the expected charge or time response of PMTs first, which has a strong dependence on the vertex or energy. 
Given the observed charge or time information of PMTs, a maximum likelihood method is then utilized to reconstruct the positron vertex or energy.
However, in the following energy reconstruction studies applied to JUNO~\cite{Wu_2019,Huang_2021}, the vertex is assumed to be known and the electronic effects of PMTs are not considered. The PMT time probability density function (PDF) of the vertex reconstruction studies in~\cite{Liu_2018, Li_2021} is vertex independent and relies on Monte Carlo simulation.
This paper developed a simultaneous reconstruction of the positron vertex and energy, using both charge and time information of PMTs, as well as the required PMT response extracted from calibration data, to improve the reconstruction precision. The major updates with respect to the previous studies are listed below:
\begin{itemize}
    \item more realistic expected charge response of PMTs with all electronic effects included 
    \item more realistic time PDF of PMT photon hits has been constructed based on $^{68}$Ge calibration data rather than positron data from Monte Carlo simulation
    \item the dependence on the propagation distance of time of flight or effective refractive index of photons in LS are calibrated, leading to a more accurate time of flight 
    \item more accurate time PDF of PMT photon hits which takes into account the dependence on the vertex radius and photon propagation distance 
    \item simultaneous reconstruction of vertex and energy with both charge and time information of PMTs
\end{itemize}

The rest of this paper is organized as follows: Sec.~\ref{sec:detector} briefly describes the JUNO detector and Monte Carlo samples. Sec.~\ref{sec:pdfs} presents the updates of the crucial inputs to the reconstruction. Sec.~\ref{sec:EVRec} describes the reconstruction method and performance. Finally, Sec.~\ref{sec:sum} gives the conclusion.


\section{JUNO detector and data samples}
\label{sec:detector}
The JUNO detector consists of the central detector (CD), the top tracker detector and the water Cherenkov detector. The target matter of the CD is 20k ton liquid scintillator filled in a 35.4 m acrylic ball, monitored by about 12612 20-inch MCP-PMTs, 5000 20-inch Dynode-PMTs and 25600 3-inch PMTs~\cite{junoppnp}. The compositions of the liquid scintillator are PPO, LAB and bis-MSB~\cite{junoLS}. 
In addition, JUNO also has a comprehensive calibration system, which consists of the Cable Loop System (CLS), the Auto Calibrate Unit (ACU), the Guide Tube Calibration System (GTCS) and the Remotely Operated under-LS Vehicles (ROV). More details about the calibration system can be found in Ref.~\cite{calib_2020}. For all the studies in this paper, only CLS and ACU will be used.

\begin{figure}[ht]
\centering
    \includegraphics[width=\figwb\textwidth]{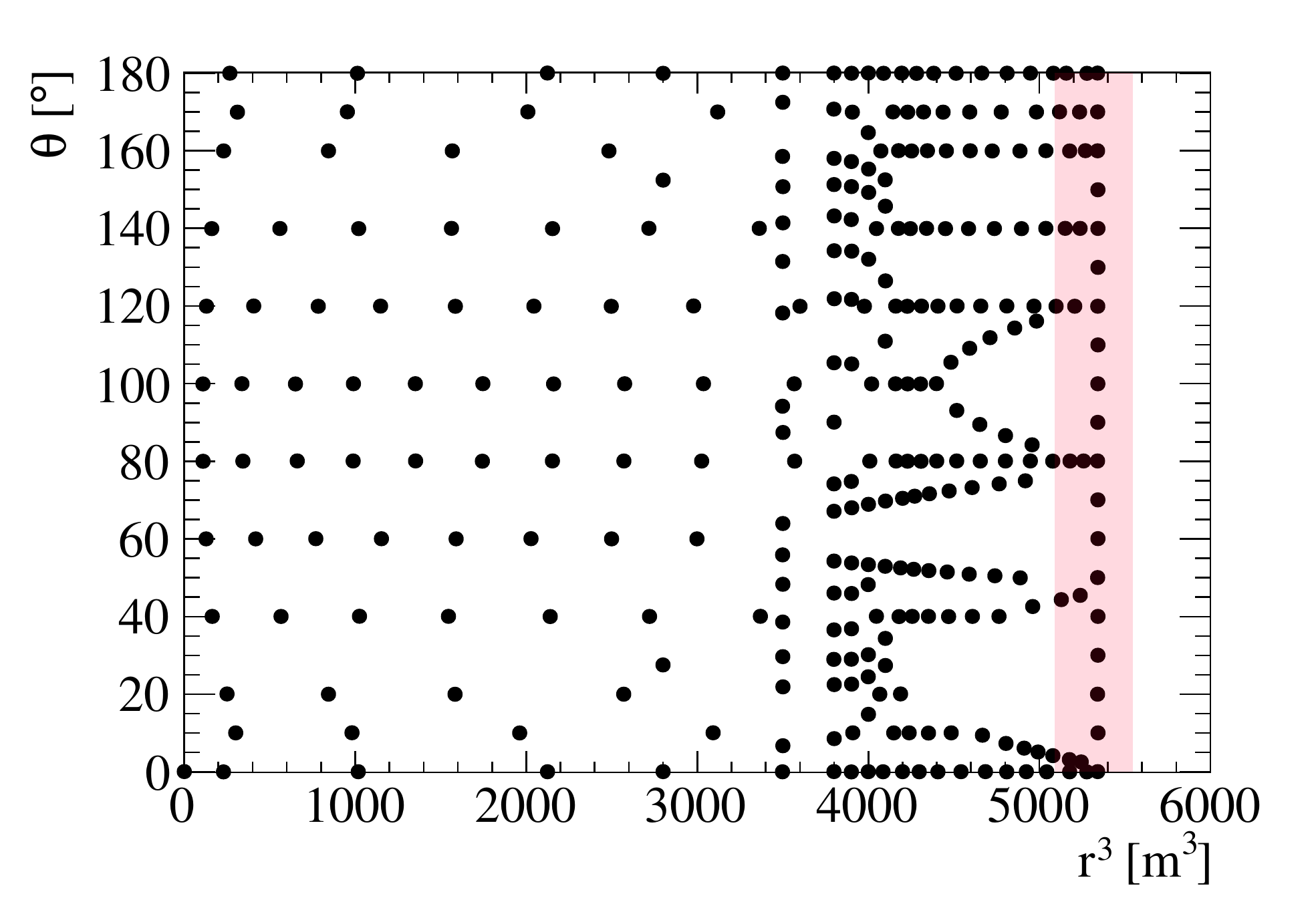}
\caption{Calibration positions on the X-Z plane of CD~\cite{Huang_2021}.}
\label{fig:calibpos}
\end{figure}

Since the JUNO detector is still under construction, Monte Carlo samples are simulated using a  custom Geant4-based (version 4.10.p02) offline software SNiPER~\cite{sniper}. 
The information of the calibration and physics data samples is summarized in Tab.~\ref{tab:datasample}.  
Laser and $^{68}$Ge calibration samples are used to construct the crucial inputs for the vertex and energy reconstruction. Calibration positions (Fig.~\ref{fig:calibpos}) are set as Case 5 described in Ref.~\cite{Huang_2021} and 10k events are simulated for each position. Nine sets of positron samples with discrete kinetic energy $E_k$ = (0, 1, 2, ... , 8) MeV are produced to evaluate the reconstruction performance. The statistics for each set is 450k and events are uniformly distributed in the CD. Electron data is generated for elaborating the construction principle and performance of the time PDF of positrons. 

\begin{table}[ht]
\centering
\caption{MC data samples. "op" stands for optical photons. "Pos." stands for position. "Energy" is deposited energy.}
\begin{tabular}{ccccc}
\toprule
Source&	 Type&	Energy [MeV]&	Pos.&	Stats. \\
\hline
Laser &	 op&	$\sim1$&	296&	10k/pos. \\
$^{68}$Ge  &	$\gamma$&	1.022& 	296&	10k/pos. \\
positron&	$e^+$&	(0,1,2,…,8)+1.022&	uniform&	450k/energy \\
electron &	$e^-$& 1&	296&	10k/pos. \\
\bottomrule
\end{tabular}
\label{tab:datasample}
\end{table}

For all these samples, realistic detector geometry is deployed. The optical parameters of LS based on measurements~\cite{junoLS} are implemented as well. Various optical processes, including scintillation, Cherenkov process, absorption and re-emission, Rayleigh scattering and reflection or refraction at detector boundaries are simulated with Geant4 in the detector simulation. 
In addition, the electronic effects of PMTs such as charge smearing, transit time spread and dark noise (DN) are implemented by a toy electronic simulation. The PMT parameters are taken from PMT testing~\cite{pmtTest, Li_2021} and are summarized in Tab.~\ref{tab:ElecPara}. These values are still being refined and electronics testing is ongoing~\cite{newPmtTest}. Although the parameters are different PMT by PMT, they roughly follow Gaussian distributions for each type of PMT. One exception is the dark noise rate, which has a much wider and non-symmetric spread. 

\begin{table}[ht]
\centering
\caption{Electronic parameters of 20-inch PMTs.}
\begin{tabular}{ccc}
\toprule
 &	 Dynode-PMT&	MCP-PMT \\
\hline
Charge resolution &	 $ 0.28 \pm 0.02$ p.e. &  $0.33 \pm 0.03$ p.e.\\
Time transit spread &   $1.1 \pm 0.1$ ns&	$7.6 \pm 0.1 $ ns \\
Dark noise  & 	$15 \pm 6 $ kHz&	$32 \pm 16$ kHz \\
\bottomrule
\end{tabular}
\label{tab:ElecPara}
\end{table}

\begin{figure}[ht]
\centering
    \includegraphics[width=\figwb\textwidth]{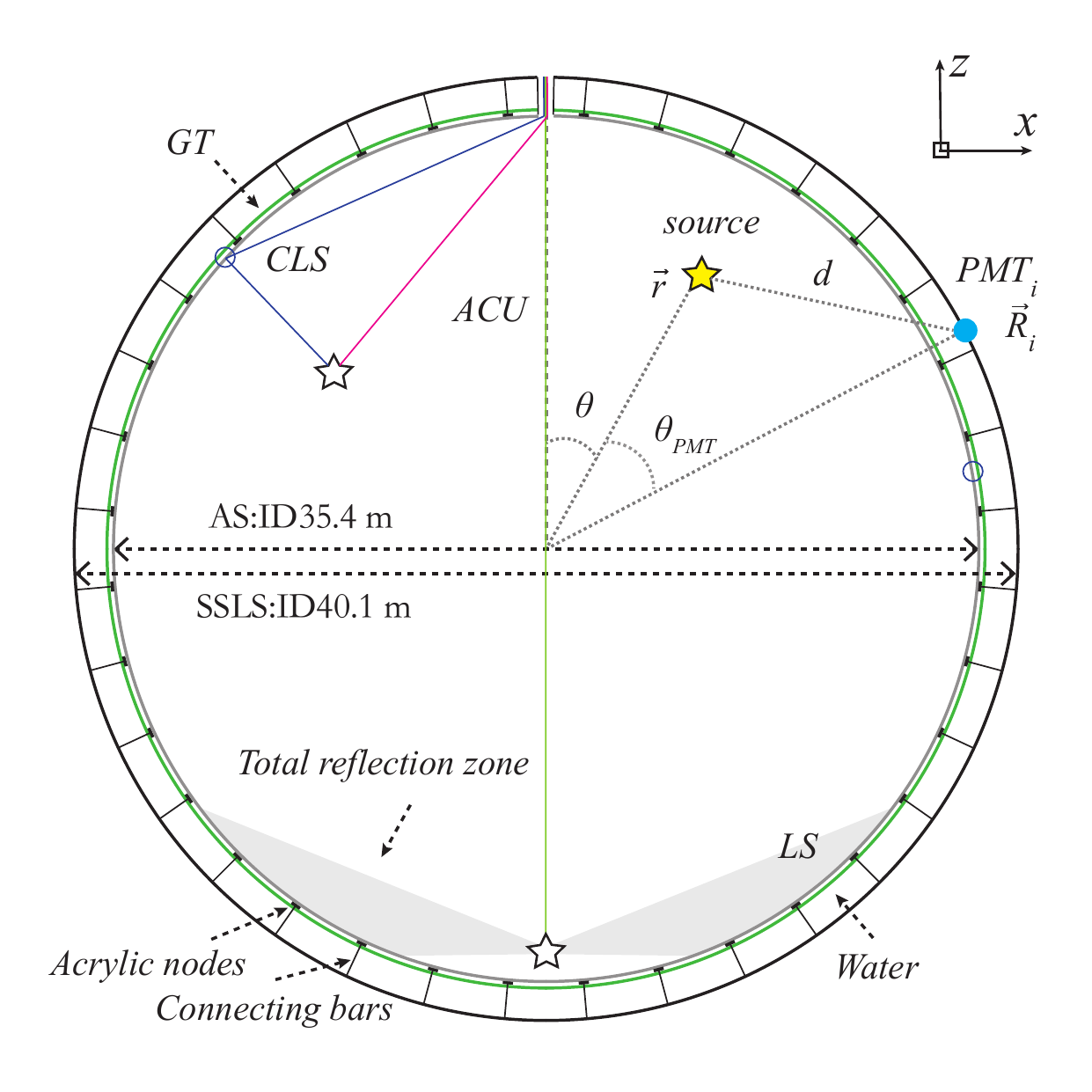}
\caption{Illustration of the variables of nPE map.}
\label{fig:MapPara}
\end{figure}

\section{Construction of the nPE map and time PDF of PMTs using calibration data}
\label{sec:pdfs}
The basic strategy of energy or vertex reconstruction is similar to that in Ref.~\cite{Huang_2021,Li_2021}. For any positron event, 
the charge and time response of all the PMTs strongly depend on the positron vertex and energy.
Firstly the expected charge (referred to as the nPE map) and time PDF of PMTs are constructed using calibration data. Given the observed charge and time information of PMTs, a likelihood function is built and then utilized to reconstruct the energy or vertex. 
As mentioned in the introduction, with respect to Ref.~\cite{Huang_2021,Li_2021}, a few important updates regarding the expected charge and time response of PMTs will be implemented in this paper and the details will be described in this section.

\subsection{Realistic nPE map with full electronic effects}
 One of the crucial components of energy reconstruction in Ref.~\cite{Huang_2021} is the nPE map denoted as $\hat{\mu}^L(r,\theta, \theta_{PMT})$. It describes the expected number of LS photoelectrons per unit visible energy. The visible energy is defined as $PE_{total}/Y_0$, where $PE_{total}$ is the total number of PEs, $Y_0$ is the constant light yield defined in Ref.~\cite{calib_2020}. The definition of $r,\theta, \theta_{PMT}$ are illustrated in Fig.~\ref{fig:MapPara}. A data-driven method of constructing the nPE map has been introduced in~\cite{Huang_2021}, where the electronic effects are omitted. 
 In a realistic case with full electronic effects, the observable of PMTs will change from nPE to charge due to the charge smearing of every single photoelectron. Meanwhile, dark noise will also contribute to the total charge of each PMT. Consequently, the formula of $\hat{\mu}^L$ has to be modified accordingly after taking these two effects into account. The updated formula is given by Eqn.~\ref{eq:EqMu-Hat},  
\begin{equation}
    \hat{\mu}^L(r, \theta, \theta_{PMT}) = \frac{1}{E_{\textrm{vis}}} \frac{1}{M}   \sum_{i=1}^{M}  \frac{\frac{\bar{q_{i}}}{\hat{Q_{i}}}-\mu_{i}^{D}} {DE_{i}}, 
    \mu_{i}^{D} = DNR_{i}\times L,
\label{eq:EqMu-Hat}
\end{equation}
where $i$ runs over PMTs with the same $\theta_{PMT}$, $DE_{i}$ is the detection efficiency. 
Photoelectrons from LS have a strong temporal correlation while dark noise photoelectrons occur randomly in time. The length of full electronic readout window is $L_{\textrm{FADC}}$ =1250 ns.
To reduce the impact of dark noise, a signal window $[t^A_{r}, t^B_{r}]$ is set according to the residual time distribution (see Eqn.~\ref{eq:time3}) of positrons to exclude most of the dark noise. 
Those dark noise photoelectrons within the signal window will contaminate the photoelectrons from the physics signal, thus the expected number of dark noise photoelectrons $\mu_{i}^{D}$ needs to be subtracted. Here $\mu_{i}^{D}$ is proportional to the dark noise rate $DNR_{i}$ and the signal window length $L=t^B_r-t^A_r$, which is optimized as 280 ns by scanning the value in [160 ns, 540 ns] and picking the one with the best effective energy resolution~\cite{juno} in this study. 
On the other hand, the average detected nPE $\Bar{n_i}$ will now be estimated with $\frac{\bar{q_{i}}}{\hat{Q_{i}}}$, where $\bar{q_{i}}$ is the average recorded charge inside the signal window and $\hat{Q_{i}}$ is the expected average charge of 1 photoelectron. 
Besides the two changes in Eqn.~\ref{eq:EqMu-Hat}, the construction procedure of the nPE map is the same as Ref.~\cite{Huang_2021}.

\subsection{Calibration of the effective refractive index of photons in LS}
\begin{figure}[h]
\centering
    \includegraphics[width=\figwb\textwidth]{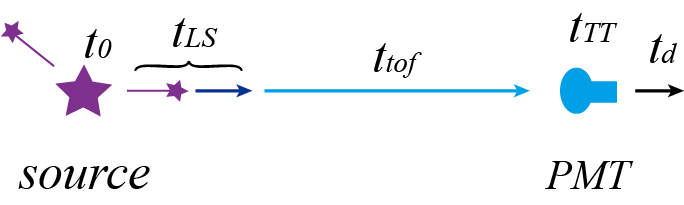}
\caption{Illustration of the components of photon hit time.}
\label{fig:timecomp}
\end{figure}

In addition to charge, the other important observable of PMTs is the hit time of photons. It can be approximately expressed as Eqn.~\ref{eq:time1}:
\begin{equation}
\label{eq:time1}
    t_h^{\prime} = t_0 + t_{LS} + t_{tof} + t_{TT}^{\prime} + t_d,
\end{equation}
 and the different components are shown in Fig.~\ref{fig:timecomp} with a simple illustration, where $t_0$ is the event starting time, $t_{LS}$ is the scintillation time which is governed by the LS optical properties, $t_{tof}$ is the time of flight of photons propagating from the event vertex to PMTs, $t_{TT}^{\prime}$ is the transit time of PMTs which roughly obeys a Gaussian distribution with ($\mu_{TT}$, $\sigma_{TT}$), $t_d$ is the delay time caused by the PMT readout electronics. Both $\mu_{TT}$ and $t_d$ are PMT dependent, and the difference among PMTs can be calibrated and absorbed by $t_h^{\prime}$, resulting the calibrated hit time $t_h$:
\begin{equation}
\label{eq:time2}
    t_h = t_0 + t_{LS} + t_{tof} + t_{TT},
\end{equation}
where $t_{TT}$ is described by the new Gaussian ($\overline{\mu_{TT}+t_d}$, $\sigma_{TT}$) and $\overline{\mu_{TT}+t_d}$ is the average value of all PMTs.

\begin{figure}[!ht]
\centering
    \includegraphics[width=\figwb\textwidth]{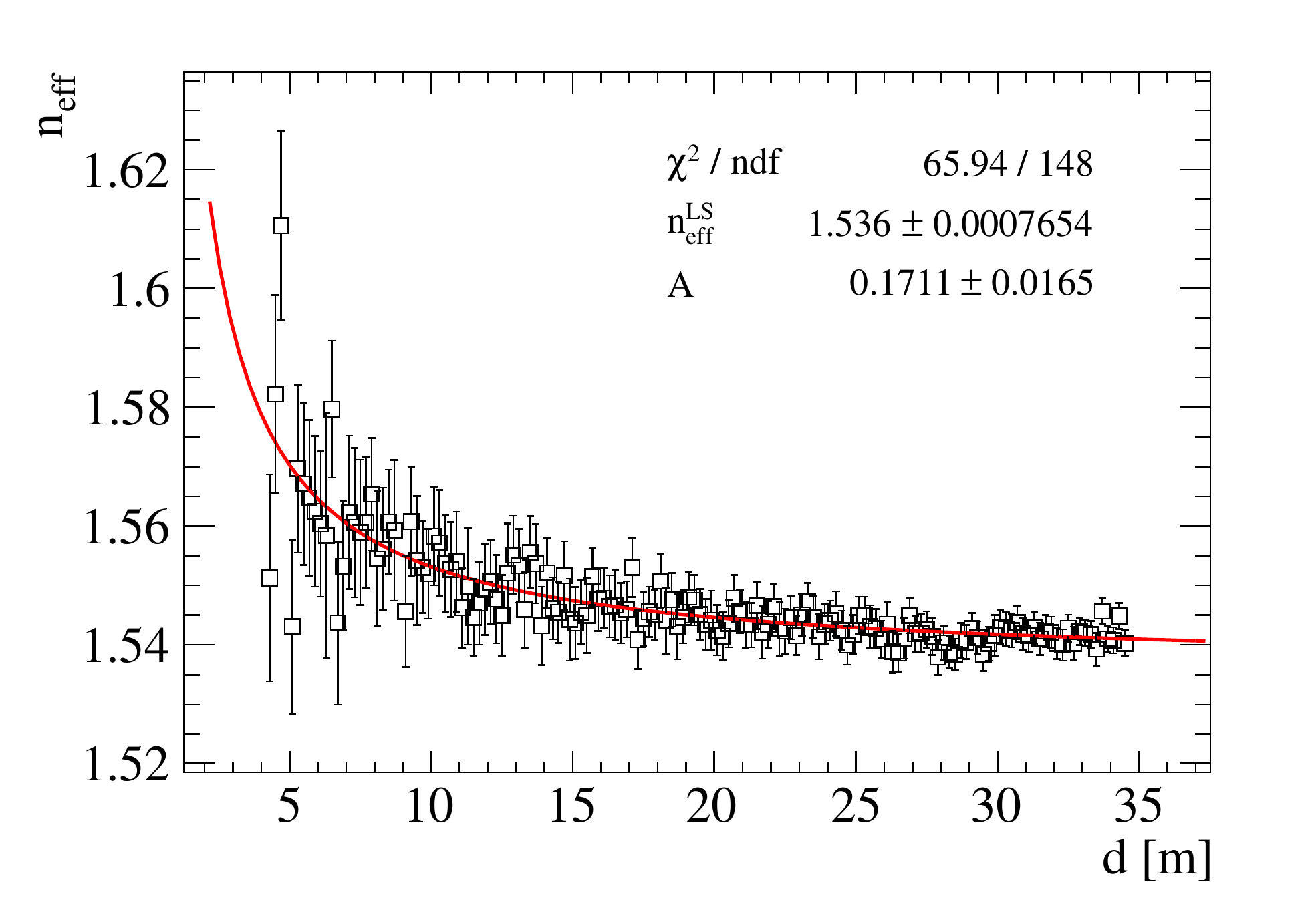}
\caption{The fitting results of effective refraction index. The uncertainty of $t_{peak}$ is set as 0.25 ns. For small values of d, the source is located near the north or south pole and the corresponding PMTs fall in the dark zone due to total reflection.}
\label{fig:RIsfit}
\end{figure}

Ref.~\cite{Li_2021} defined the residual time of PMT photon hits $t_r$ as 
\begin{equation}
\label{eq:time3}
    t_r = t_h - t_{tof} - t_0 \\
    = t_{LS} + t_{TT}.
\end{equation}
To calculate the time of flight of photons in JUNO CD, a constant effective refractive index was used in Ref.~\cite{Li_2021}. 
However, the propagation of photons in LS is complicated and includes various optical processes as mentioned in Sec.~\ref{sec:detector}.
The time of flight might not necessarily be proportional to the propagation distance.  Given the calibration sources could be deployed at different positions in the CD, one could potentially use the calibration data to calibrate the time of flight or equivalently the effective refractive index $n_{eff}$ as a function of the propagation distance. 
For radioactive sources, the starting time $t_0$ of each event is unknown. On the other hand, the precision of $t_0$ can reach $<$ 0.5~ns level for the Laser source~\cite{junoLaser}.
Thus Laser source is chosen to calibrate $n_{eff}$.
Although the original optical photons of the Laser source have a fixed wavelength, they will be quickly absorbed and re-emitted by LS, resulting a wavelength spectrum  similar to that of other sources~\cite{junoLaser}.  
By deploying the Laser source at different positions along the z-axis of the CD with ACU, one can plot the distribution of $t_r + t_{tof} = t_h - t_0$ for each PMT with enough statistics of Laser events. 
Since $t_{LS}$ only depends on the LS properties, it is the same for all PMTs. Meanwhile, $t_{TT}$ defined in Eqn.~\ref{eq:time3} roughly follows the same distribution for the same type of PMTs. 
Thus the shape of the $ t_h - t_0$ distribution should be the same for the same type of PMTs and the $t_{tof}$ term merely leads to a relative shift with respect to $t_r$, which could be measured by the peak of the $t_h - t_0$ distribution
\begin{equation}
\label{eq:time4}
    t_{peak} = t_{tof} + t_{peak}^0,
\end{equation}
where $t_{peak}^0$ is the peak of the $t_r$ distribution and it is a constant with different values for each type of PMTs. 
Let us define the distance between PMT and the source position as d (Fig.~\ref{fig:MapPara}). For PMTs with the same type and d, their $t_{peak}$ should be aligned, since $t_{tof}$ is the same at first order approximation. Thus their $t_{peak}$ is averaged to reduce any small second order fluctuations.
Given that Dynode PMTs have much better time resolution than MCP PMTs, only the Dynode PMTs are used to obtain more precise $t_h$. 
By deploying the Laser source at different positions, one could obtain a large set of ($t_{peak}$, d) data points, which could then be used to fit $t_{tof}$ as a function of d. 
For convenience the effective refraction index $n_{eff}(d)$ is used instead of $t_{tof}(d)$:  
\begin{equation}
\label{eq:effIndex}
     n_{eff}(d) = \frac{c\times (t_{peak}(d) - t_{peak}^0)} {d}.
\end{equation}
The fitting function of the effective refraction index is given by 
\begin{equation}
\label{eq:EqNeffFunc}
     n_{eff} = n^{LS}_{eff}  + \frac{A}{d},
\end{equation}
 
\noindent where $n^{LS}_{eff}$ is the dominant term of the effective refraction index of liquid scintillator, $A$ is the coefficient of the additional term which is distance dependent. 
Since $t_{peak}^0$ is $\sim 1$ ns and nontrivial to calibrate, it is conservatively set to 0, which modifies $A$ term and has little impact on the reconstruction performances. Fig.~\ref{fig:RIsfit} shows the fitting results. The best fit value $n^{LS}_{eff}\approx 1.54$ is consistent with Ref.~\cite{Li_2021}.

\subsection{Construction of more accurate and realistic time PDF.}
\noindent One caveat of the residual time PDF from Ref.~\cite{Li_2021} is that it was obtained from MC simulation. Any potential discrepancy between MC and real data will degrade the vertex reconstruction. 
Furthermore, the PDF was constructed only using events at the detector center. This simplification did not consider the PDF dependence on the vertex and could potentially explain the large vertex bias observed near the detector border in Ref.~\cite{Li_2021}. Inspired by the calibration data-driven construction of the nPE map in Ref.~\cite{Huang_2021}, one could also use the same calibration data to build a realistic time PDF. Moreover, the various calibration positions allow for a more precise parametrization of the time PDF.

For the construction of the nPE map in Ref.~\cite{Huang_2021}, $^{68}$Ge and Laser sources were used. As mentioned previously, the optical photons of the Laser source will be absorbed and re-emitted by LS, 
however other particles have to transfer energy to the LS molecules first. 
As a result, the photon timing profile of the Laser source is different with respect to that of positrons. In this paper, we will only use $^{68}$Ge to construct the time PDF of positrons. 
Ideally, the electron source could be used to describe the time PDF of photons originating from the kinetic energy part of positrons.
Although electron sources are not available, we compared the reconstruction performance of using a time PDF of $^{68}$Ge to that of $^{68}$Ge and electrons, to check its impact. The details will be discussed in Sec.~\ref{sec:EVRec}. 

As shown in Eqn.~\ref{eq:time3}, in order to calculate the residual time, the information of $t_0$ is needed for every single event.
Different from the laser source with high precision of $t_0$, this quantity of calibration events from radioactive sources is unknown. The reconstruction of $t_0$ for each event was attempted, but the uncertainty was larger than 2~ns. 
Since the $t_h - t_{tof}$ distribution for different events with fixed vertex and energy should roughly have the same shape, the different $t_0$ would merely cause a relative shift of the distribution. 
One could use the peak of  $t_h - t_{tof}$ distribution $t^{'}_0$ as the new reference time instead of $t_0$ to align different events. 
Eqn.~\ref{eq:time3} can be modified as below so that the distribution of the newly defined residual time $t_r^{'}$ always peaks at 0. For convenience, the prime symbol of $t_r^{'}$ and $t_0^{'}$ will be omitted throughout the rest of the paper.
\begin{equation}
\label{eq:time5}
    t_r^{'} = t_h - t_{tof} - t^{'}_0 \\
    = t_{LS} + t_{TT} + constant.
\end{equation}

The PDF of $P_T(t_r|r, d, \mu_l, \mu_d,k)$ describes the probability of the residual time of the first photon hit falling in $[t_r, t_r+\delta t]$: $r$ is the radius of the event vertex,  $d=|\vec{r} - \vec{r}_{PMT}|$ is the propagation distance from the previous section, $\mu_l$ and $\mu_d$ are the expected number of LS and dark noise photoelectrons inside the full electronic readout window respectively, $k$ is the detected number of photoelectrons. 
The dependence of $P_T$ on the parameters will be discussed in the following sub-sections. For convenience, let us denote the $f(t)$ as the probability density of ``photoelectron hit at time t" and $F(t)$ as the probability of ``photoelectron hit after time t" for a general case. Then $F(t)$ can be calculated by
\begin{equation}
\label{eq:LSpdfElem}
     F(t) = \int_{t}^{L_{FADC}} f(t')dt'.
\end{equation}

\subsubsection{Dependence on r and d}
Optical processes such as absorption and re-emission or Rayleigh scattering are not negligible in a LS volume as large as JUNO, and they become more prominent as the photon propagation distance d increases. 
Meanwhile, total reflection could significantly change the direction of photons, which is more likely to happen for events with larger radius r. Thus the $f(t)$ of LS photoelectron $f_l(t)$ is  dependent on d and r. 
This dependency can be addressed by deploying calibration sources at different positions with the ACU + CLS system.
This data-driven construction of time PDF does not require a comprehensive understanding of the properties of the liquid scintillator such as the attenuation length and decay time.

\begin{figure*}[ht]
\centering
    \subfigure[ $ 0~m^3 < r^3 < 1000~m^3 $ ]  {\includegraphics[width=\figwb\textwidth]{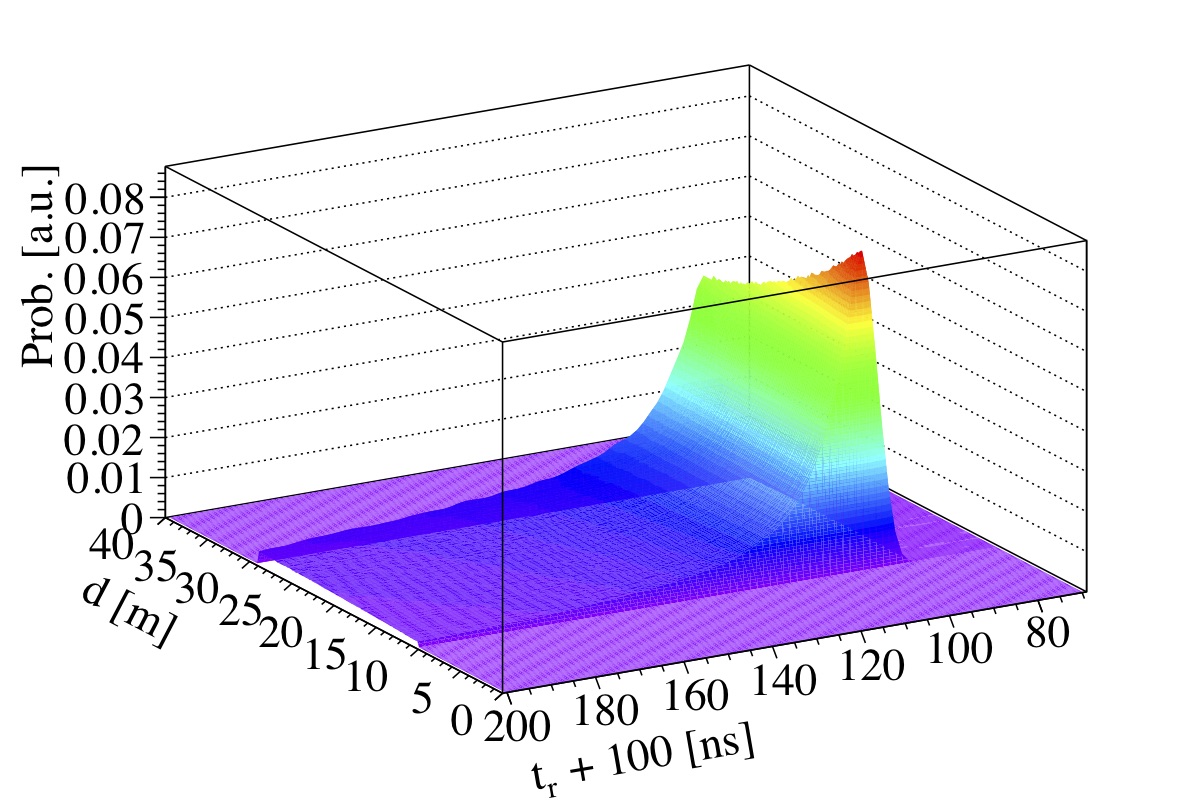}}
    \subfigure[ $ 4500~m^3 < r^3 < 4550~m^3 $ ]  {\includegraphics[width=\figwb\textwidth]{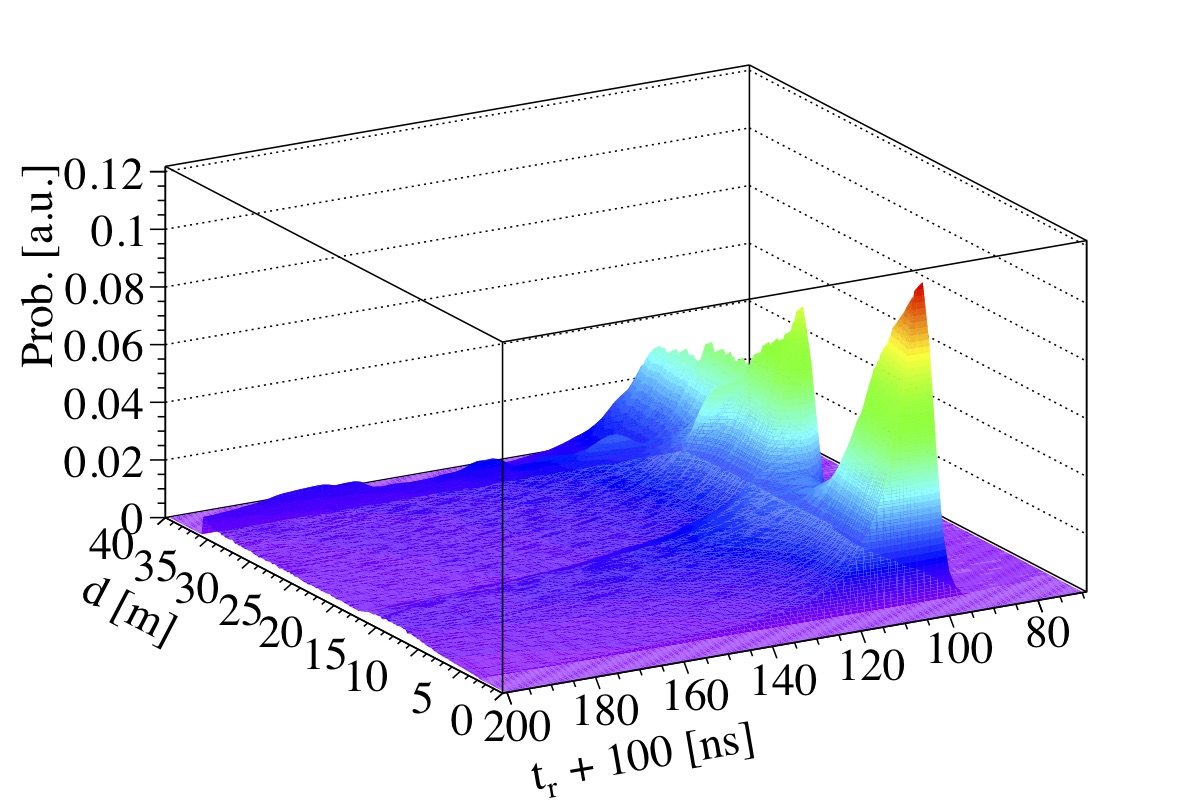}}
\caption{The shape of time PDFs at different positions. The time PDF of total reflection region is different from that of central region and they becomes wider as d increase.}
\label{fig:TimePdf2D}
\end{figure*}

One of the challenges to construct the time PDF from the calibration data is that the number of calibration positions is limited. To address this challenge, 35 bins and 200 bins are set in the r-direction and d-direction, respectively. Examples of the construction results of the $t_r$ PDF of 1 LS photoelectron are shown in Fig.~\ref{fig:TimePdf2D}. The left and right plots correspond to vertices in the central region and total reflection region respectively. 
One can clearly see the difference of the time PDF for different radius r. Also, $t_r$ PDF becomes wider as $d$ increases.
One thing to note is that the contribution from dark noise has been subtracted and will be added independently in the next subsection.

Once the time PDF of 1 LS photoelectron is obtained, it is straightforward to calculate the time PDF of n LS photoelectrons using Eqn.~\ref{eq:nLPETpdf}, where $I^{l}_n$ is a normalization coefficient.
\begin{equation}
\label{eq:nLPETpdf}
      P^l_T(t|k=n) = I_n^l\times[f_{l}(t)F^{n-1}_l(t)].
\end{equation}

\subsubsection{Adding dark noise contribution}
Photoelectrons induced by PMT DN will contaminate the photoelectrons from the signal particles in LS. Their impact on the residual time PDF must be taken into account carefully.
Since dark noise photoelectrons occur randomly in time, their $f(t)$ is simply $f_d (t)=1/L_{\textrm{FADC}}$. 
The probability of a DN photoelectron falls in $[t, L_{FADC}]$ is given by Eqn.~\ref{eq:DNpdfElem}.
\begin{equation}
\label{eq:DNpdfElem}
     F_{d}(t)  = 1 - \frac{t}{L_{FADC}}.
\end{equation}

\begin{figure}[ht]
\centering
    \subfigure[ k = 1 p.e.. ]  {\includegraphics[width=\figwb\textwidth]{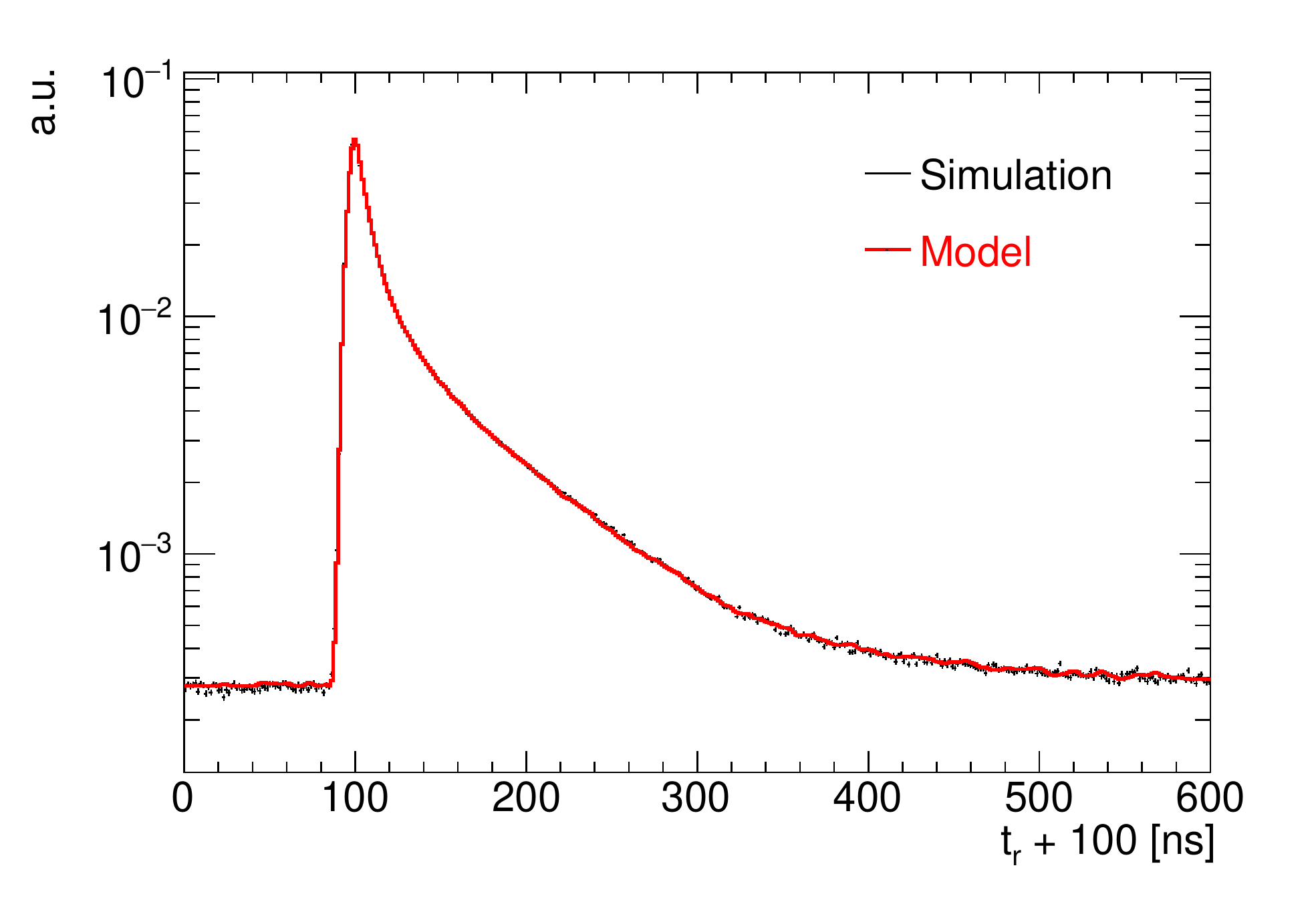}}
    \subfigure[ k = 2 p.e.. ]  {\includegraphics[width=\figwb\textwidth]{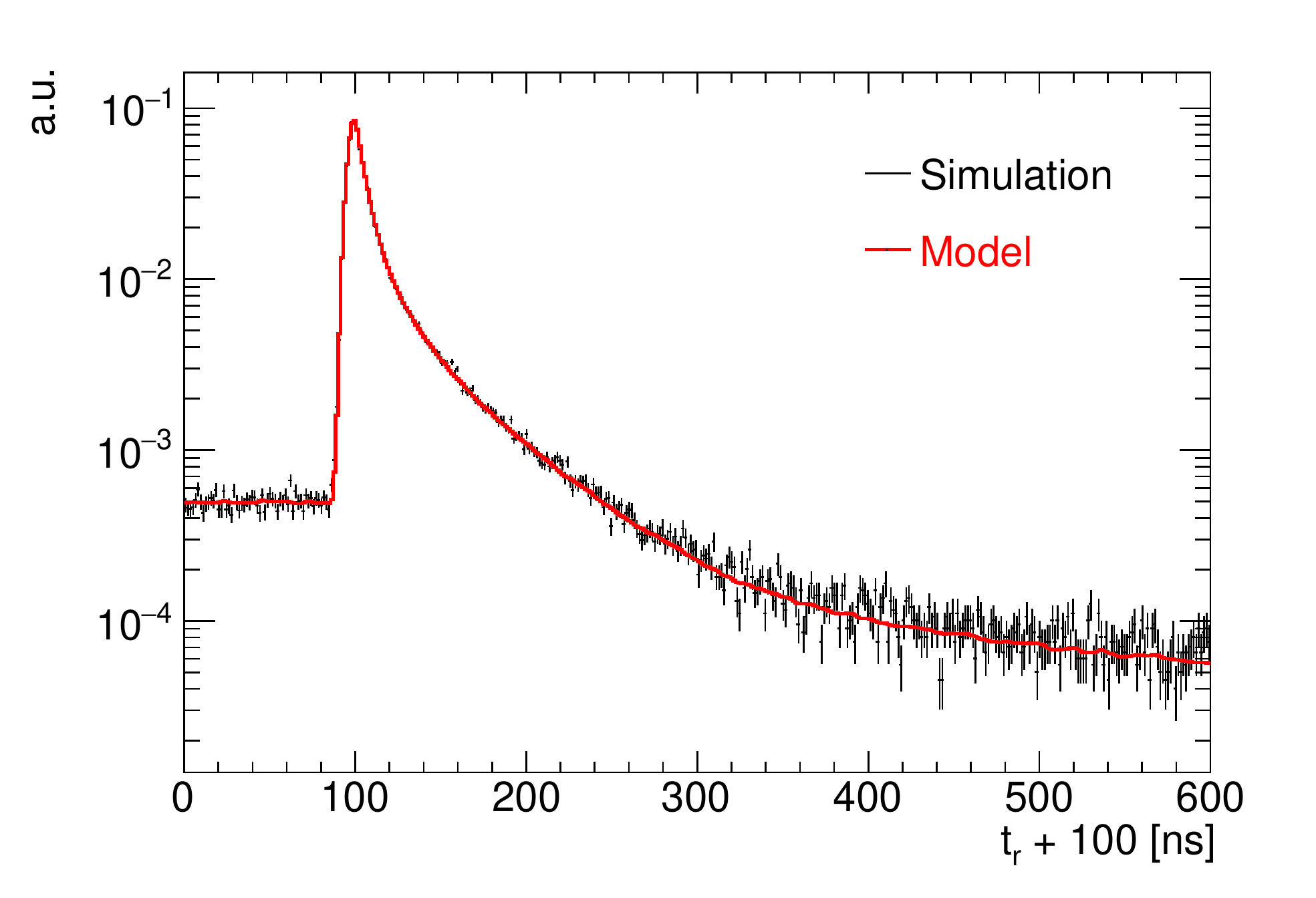}}
 
\caption{Time PDF of Dynode-PMT of central $^{68}$Ge events considering dark noise. Red histograms are calculated by Eqn.~\ref{eq:nPETpdf} while black histograms are obtained from MC simulation data. }
\label{fig:DNTimePdf}
\end{figure}

For any given PMT with expected nPE from LS $\mu_l$, expected nPE from DN $\mu_d$ and detected nPE k, one could mathematically calculate 
the probability density of the first photon hit being observed at time $t$. In the simplest case where only 1 PE is detected by the PMT or k=1, $P_{T}(t|\mu_l, \mu_d, k=1)$ is given by
\begin{equation}\small
\label{eq:1PETpdf}
\begin{split}
    &P_{T}(t|k=1) = I_{1}\times[ P(1,\mu_l)P(0,\mu_d)f_l (t) \\
      &+ P(0,\mu_l)P(1,\mu_d)f_d(t)],
\end{split}
\end{equation}
where $I_1$ is a normalization factor, $P(k_l,\mu_l)$ and $P(k_d,\mu_d)$ are the Poisson probability of detecting $k_l$ photoelectrons from LS and $k_d$ photoelectrons from DN respectively, 
with the condition $k$=$k_l$+ $k_d$. Given $k$=1 in this case, both $k_l$ and $k_d$ can be either 1 or 0. Thus one can easily see that the two terms correspond to the photoelectron coming from LS or DN respectively.
The same method could be applied to the case with k$\geq$2.  
The time PDF of first hit of n photoelectrons is given by
\begin{equation}\small
\label{eq:nPETpdf}
    \begin{split}
        &P_T(t|k=n) \approx I_{n}\times[ P(n,\mu_l)P(0,\mu_d))P^{l}_{T}(t|n) \\ 
     &+ P(n-1,\mu_l)P(1,\mu_d) ( P^{l}_{T}(t|n-1)F_d(t)  
        + f_d(t)F^{n-1}_l(t) ) \\
     &+ P(n-2,\mu_l)P(2,\mu_d)( P^{l}_{T}(t|n-2)F^{2}_d(t)\\
        &+ 2f_d(t)F_d(t)F^{n-2}_l(t) ) ],
    \end{split}
\end{equation}

where $I_n$ again is a normalization coefficient. The three terms correspond to the cases with $k_d$=0, 1, 2 respectively. 
Given $\mu_d$ is rather small, those terms with $k_d>2$ will be highly suppressed by the Poisson probability $P(k_d,\mu_d)$ and thus can be safely omitted to simplify the calculation. 
In order to verify this analytical approach, the calculated results of the time PDF containing the dark noise 
contributions are compared to those obtained from using the truth information of k in the MC simulation as shown in Fig.~\ref{fig:DNTimePdf}. The good agreement  
verifies the correctness of Eqn.~\ref{eq:1PETpdf} and Eqn.~\ref{eq:nPETpdf}.

\subsubsection{Charge vs. nPE}
As described in the above two subsections, the usage of time PDF requires knowledge of the number of photoelectrons detected by each PMT, whereas usually the PMT charge is measured.
Given that the charge information has been used to estimate $\mu$, the time PDF can be rewritten as Eqn.~\ref{eq:DrvMuTimePdf}:
\begin{equation}\small
\label{eq:DrvMuTimePdf}
     P_{T}^{'}(t|r, d, \mu_{l}, \mu_{d}) = \frac{\sum_{k} P_{T}(t|r, d, \mu_{l}, \mu_{d}, k)\times P(k, \mu_{l}+\mu_{d})} {\sum_{k} P(k, \mu_{l}+\mu_{d})}.
\end{equation}

\section{Simultaneous reconstruction of vertex and energy}
\label{sec:EVRec}
In previous studies~\cite{Huang_2021, Li_2021}, the reconstruction of the positron vertex and energy were decoupled for simplicity. The positron energy was reconstructed assuming its vertex is known and vice versa. 
For real data, neither the vertex nor the energy of the positron is known, they both have to be reconstructed. Meanwhile, these two variables are highly correlated. 
One main correlation is that the energy response for mono-energetic positrons varies at different vertices, which is also referred to as the detector energy non-uniformity. Another is that the vertex resolution depends on the energy as well. The higher the positron energy, the smaller the vertex resolution. In this section, a simultaneous reconstruction of the positron vertex and energy for large liquid scintillator detectors will be presented. The strong correlation between vertex and energy is naturally handled. 
Moreover, the crucial inputs of the simultaneous reconstruction, namely the nPE map and time PDF of PMTs, could be obtained from calibration data and would not depend on the MC simulation. In addition, with all the updates from Sec.~\ref{sec:pdfs}, the nPE map and time PDF of PMTs are more realistic and more accurate.

The reconstruction performance is evaluated by radial bias, radial resolution, energy uniformity and energy resolution. The radial bias and resolution are defined as the mean and sigma of the Gaussian fit of the $r_{\textrm{rec}}-r_{\textrm{edep}}$ distribution, respectively.
Energy uniformity represents the consistency of the reconstructed energy of identical particles generated at different positions, which is assessed by the deviation of the average reconstructed energies of mono-energetic positrons within different small volumes ($\sim10$ $m^3$) of the detector~\cite{Huang_2021}. 
The reconstructed energy of mono-energetic positrons will be fitted with a Gaussian function $(\bar{E}_{rec}, \sigma_{E_{rec}})$. The energy resolution is then defined as $\sigma_{E_{rec}}/\bar{E}_{rec}$. The default fiducial volume condition is $r_{rec}<17.2~\textrm{m}$.

\subsection{Charge based maximum likelihood estimation}
Ref.~\cite{Huang_2021} presented the basic strategy for energy reconstruction, a likelihood function was constructed using the expected nPE $\mu = \hat{\mu}_L \times E_{vis}$ and observed charge for each PMT, by maximizing the likelihood function one could obtain the reconstructed energy. 
However the expected nPE for each PMT strongly depends on the positron vertex, as indicated by Eqn.~\ref{eq:EqMu-Hat}. In Ref.~\cite{Huang_2021} the vertex was assumed to be known, but in real data, it needs to be reconstructed as well. Thus one could simultaneously reconstruct the vertex and energy using a likelihood function similar to that in Ref.~\cite{Huang_2021}.

\begin{figure*}[htp]
\centering
    \includegraphics[width=0.97\textwidth]{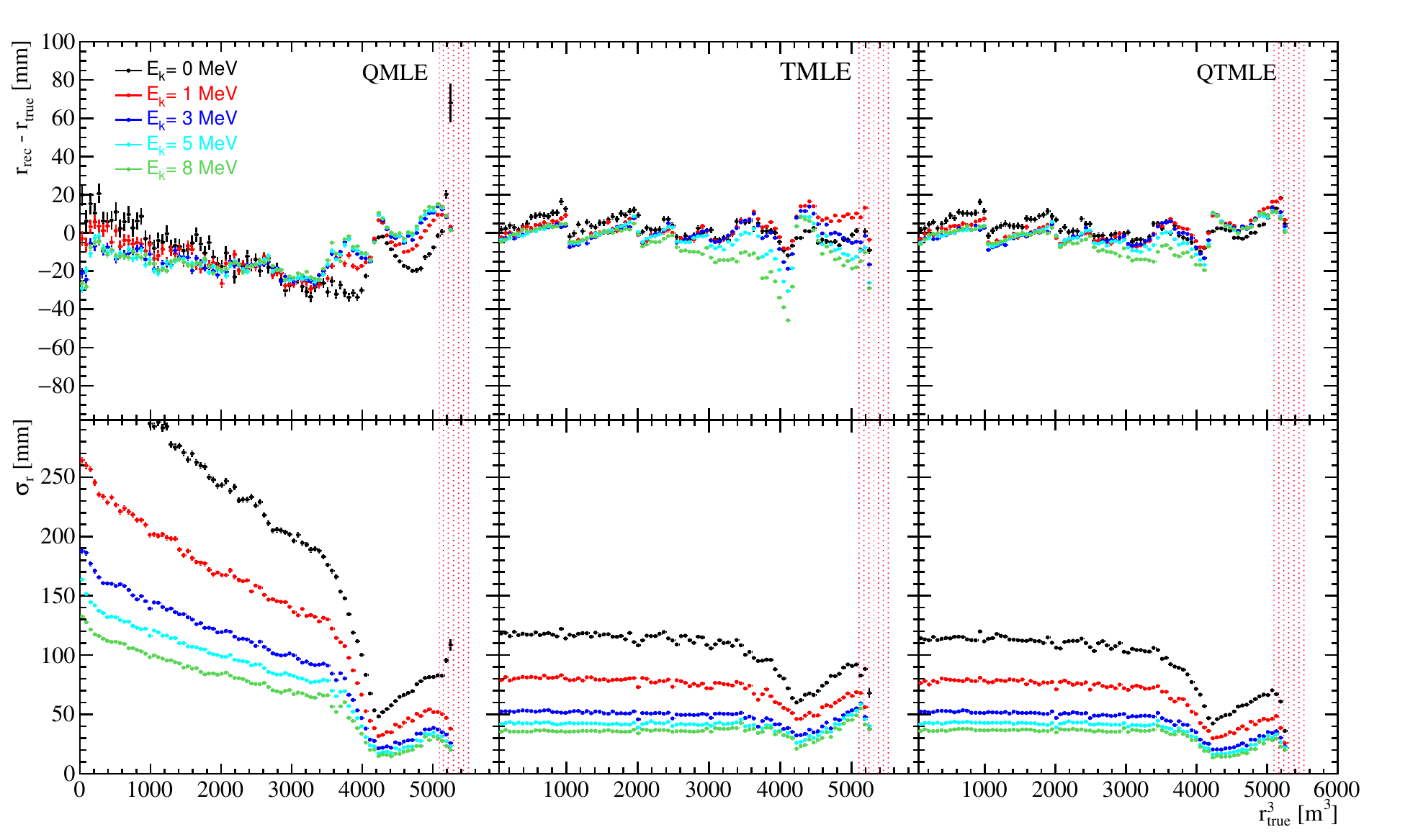}
 
\caption{Vertex reconstruction performances. The left, middle and right columns correspond to the QMLE, TMLE and QTMLE methods, respectively. The top row shows the vertex bias and the bottom row shows the vertex resolution.}
\label{fig:VRecPrfm}
\end{figure*}

This likelihood function utilizes only the charge information of PMTs and is referred to as charge-based maximum likelihood estimation (QMLE). It is constructed as Eqn.~\ref{eq:EqQMLE},  which is the product of the probabilities of observing a charge of $q_i$ when the expected nPE is $\mu_{i}$ for the $i$-th PMT. 
\begin{equation}\small
\label{eq:EqQMLE}
\begin{split}
    &\mathcal{L}(q_{1},q_{2},...,q_{N}|\mathbf{r},E_{\textrm{vis}})= \\
    & \prod_{unfired}e^{-\mu_{j}}\prod_{fired} \left(\sum_{k=1}^{+\infty} P_{Q}(q_{i}|k) \times P(k, \mu_i) \right),
\end{split}
\end{equation}
here $\mu_{i} = E_{\textrm{vis}}\times \hat{\mu}^{L}_{i} + \mu_{i}^{D}$ and $P(k, \mu_i)$ is the Poisson probability for detecting k photoelectrons. 
$P_{Q}(q_{i}|k)$ is the charge PDF of k photoelectrons that can be constructed by convolving the single photoelectron charge spectrum (SPES). 
Indices j and i run over all the "unfired" and "fired" PMTs respectively, with the PMT firing threshold of $q > 0.1~\textrm{p.e}$.
its constraint power dramatically decreases for positrons in the central region of the CD. This has been shown in Ref.~\cite{mlVertex2, Li_2021} and could also be seen in the  bottom left plot of Fig.~\ref{fig:VRecPrfm}, 
where the vertex resolution in the central region is much worse with respect to that in the border region. 
Moreover, the vertex bias of QMLE is large in the top left plot of Fig.~\ref{fig:VRecPrfm}. Inaccurate vertex will degrade the energy resolution for the simultaneous reconstruction. 

\subsection{Time based maximum likelihood estimation}

The event vertex could be strongly constrained by the time information of PMTs. Similar to Ref.~\cite{Li_2021}, a likelihood function could be constructed using the first hit time of PMTs and the more accurate and realistic time PDF $P_T$ from Eqn.~\ref{eq:DrvMuTimePdf}.
This likelihood function uses only the PMT time information and is referred to as time-based maximum likelihood estimation (TMLE). It can be constructed as Eqn.~\ref{eq:EqTMLE}, which is the product of the probabilities of observing a residual time of $t_{i,r}$ when the expected $t_{tof}$ is $d/(\textrm{c}/n_{eff})$ for the $i$-th PMT.
\begin{equation}\small
\label{eq:EqTMLE}
\begin{split}
      &\mathcal{L}(t_{1,r},t_{2,r},...,t_{N,r}|\mathbf{r}, t_0) = \\ 
      & \prod_{T-\textrm{valid}~hit} \frac{\sum^{K}_{k=1} P_{T}(t_{i,r}|r, d_i, \mu^{l}_i, \mu^{d}_i, k)\times P(k, \mu^{l}_i+\mu^{d}_i)} {\sum^{K}_{k=1} P(k, \mu^{l}_i+\mu^{d}_i)},
\end{split}
\end{equation}
here "$T-\textrm{valid}$" hit refers to those hits satisfing $-100 <t_{i,r} < 500~\textrm{ns}$ and $0.1~\textrm{p.e.} < q^{f}_{i} < K = 20~\textrm{~p.e}$.
 The definition range of the residual time PDF is (-100, 500) ns. $q^f$ stands for the total charge within the full electronic readout window. A cutoff value $K$ is set for the detected nPE $k$ to simplify the calculation.
 TMLE takes the reconstructed vertex and energy from QMLE as initial values and only updates the reconstructed vertex. Note that the reference time $t_0$ is also a free parameter in the reconstruction. As shown in Fig.~\ref{fig:VRecPrfm} and \ref{fig:VRecPrfm_res}, the vertex bias and resolution of TMLE are largely improved with respect to QMLE, especially in the central region of the CD. This is mainly due to the stronger constraint from the PMT time information.

\subsection{Charge and time combined maximum likelihood estimation}
\begin{figure*}[htp]
\centering
     \includegraphics[width=\figwb\textwidth]{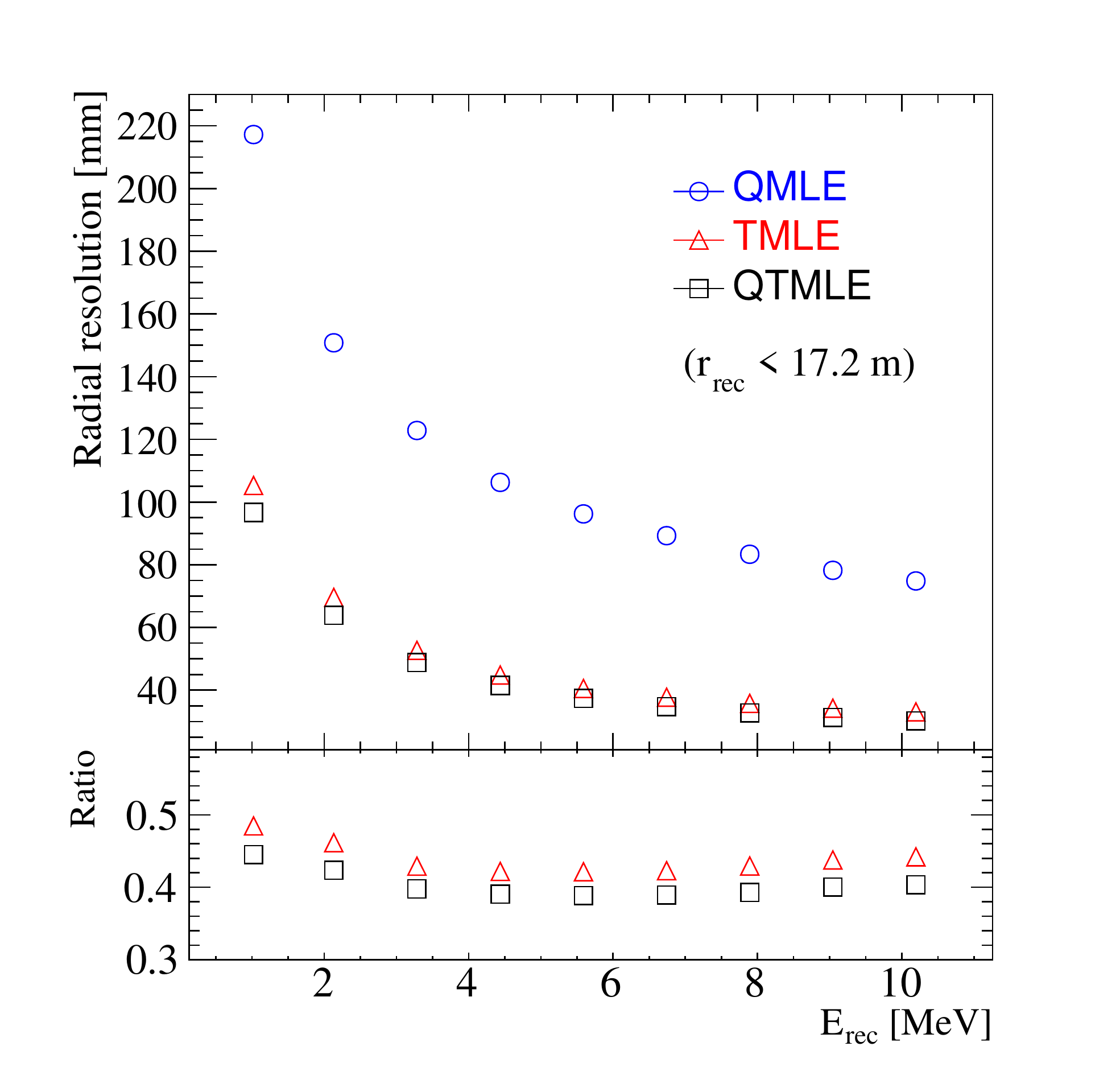}
     \includegraphics[width=\figwb\textwidth]{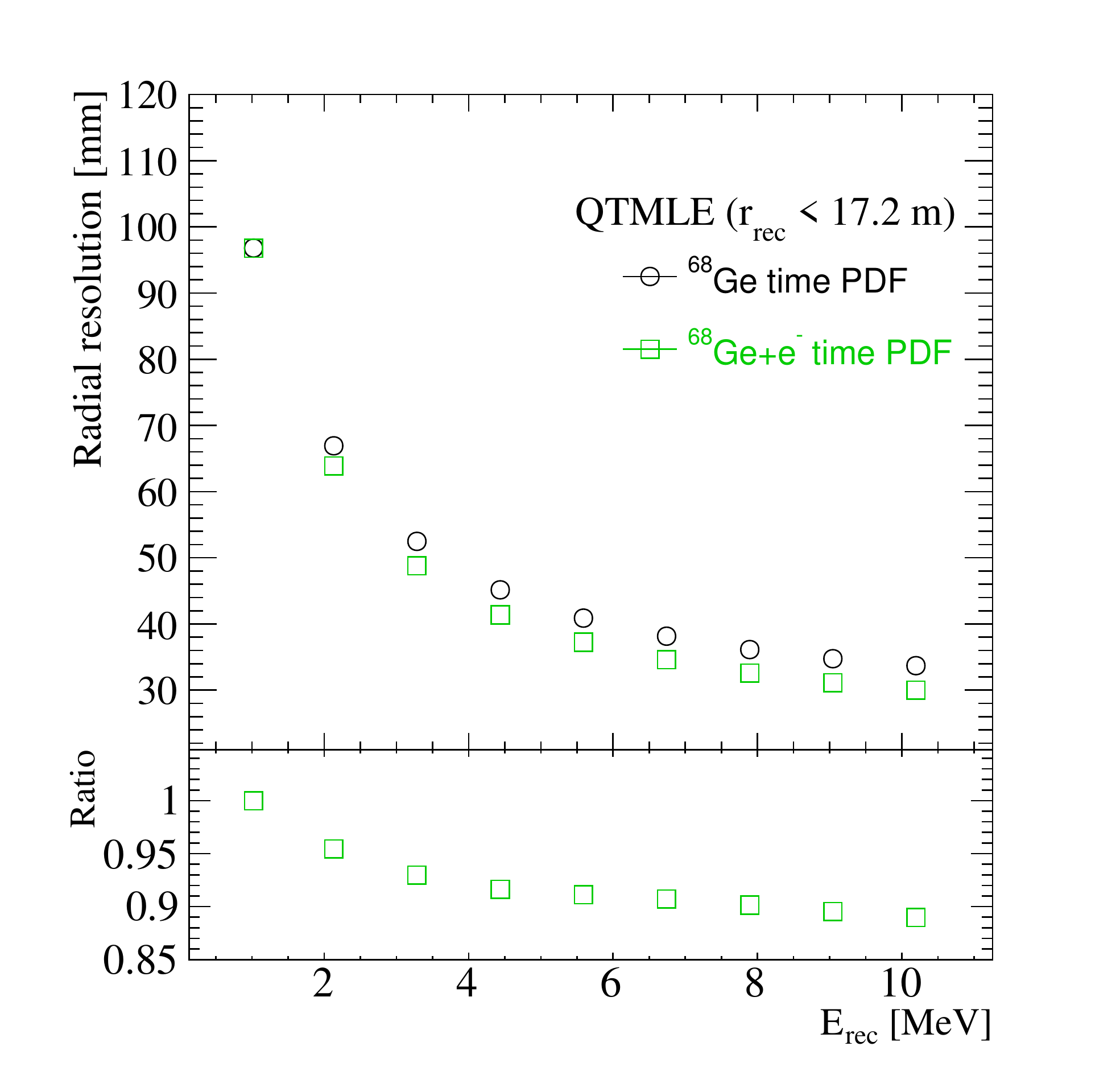}
    \caption{Comparison of the vertex resolution as a function of the energy. The left plot shows the comparison among the three methods QMLE (Blue), TMLE (Red) and QTMLE (Black). The ratio is defined as $Res_{R}(\textrm{other})/Res_{R}(\textrm{QMLE})$.
The QTMLE method utilizing both charge and time information of PMTs effectively improves the vertex reconstruction. 
    The right plot shows the comparison between using $^{68}$Ge+electron time PDF and $^{68}$Ge time PDF in the QTMLE method, which will be discussed in Sec.~\ref{sec:discussion}. The ratio is defined as $Res_{R}(^{68}\textrm{Ge}+e^-~ \textrm{time PDF})/Res_{R}(^{68}\textrm{Ge}~ \textrm{time PDF})$.}

\label{fig:VRecPrfm_res}
\end{figure*}

\begin{figure*}[!ht]
\centering
  
    \includegraphics[width=\figwb\textwidth]{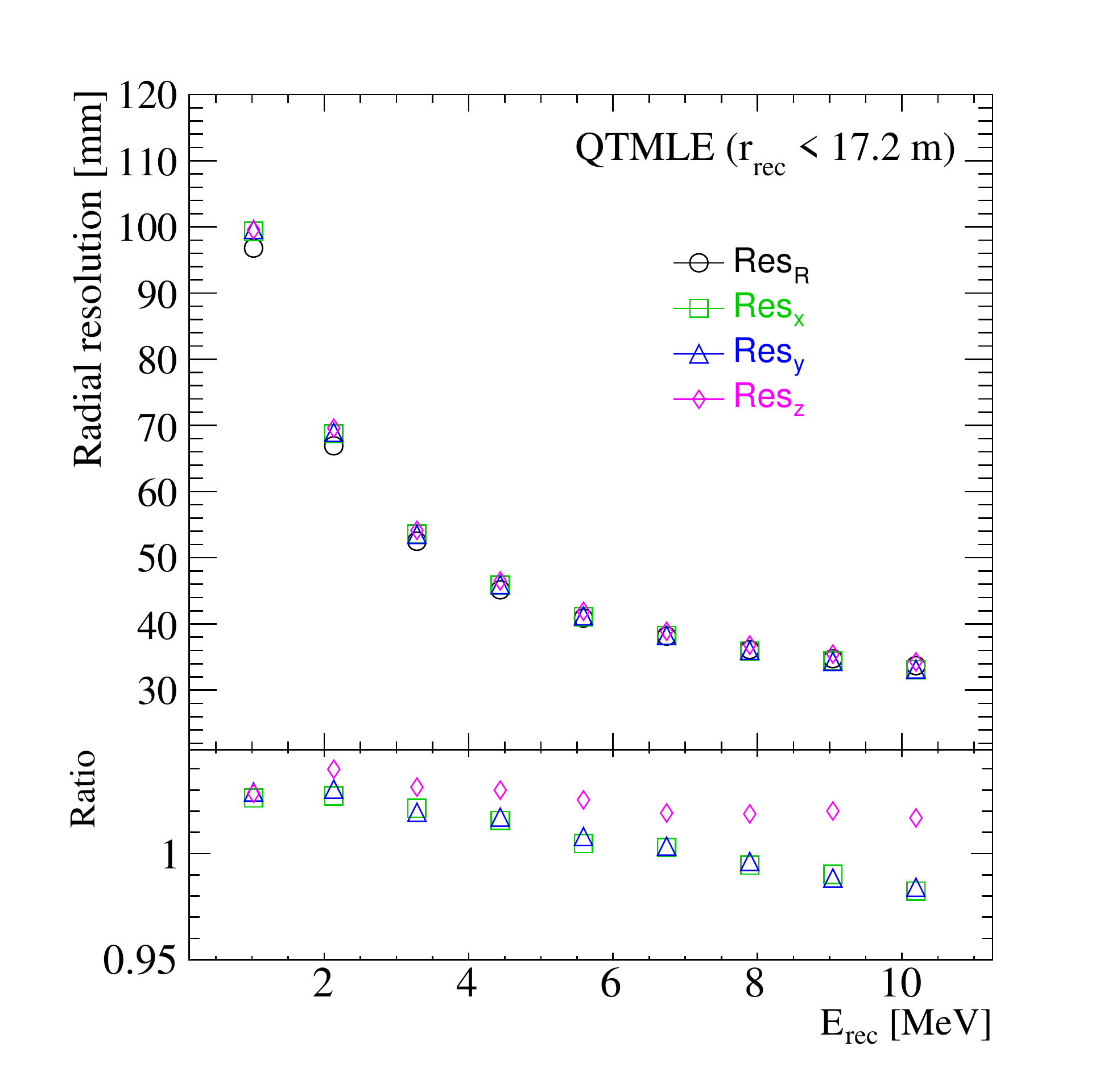}
    \caption{Comparison of the x, y, z, r resolution  of QTMLE method with $^{68}$Ge time PDF.
    The bottom panel shows the ratio to the radial resolution.}
\label{fig:VRecPrfm_XYZres}
\end{figure*}

\begin{figure*}[!ht]
\centering
  
    \includegraphics[width=\figwb\textwidth]{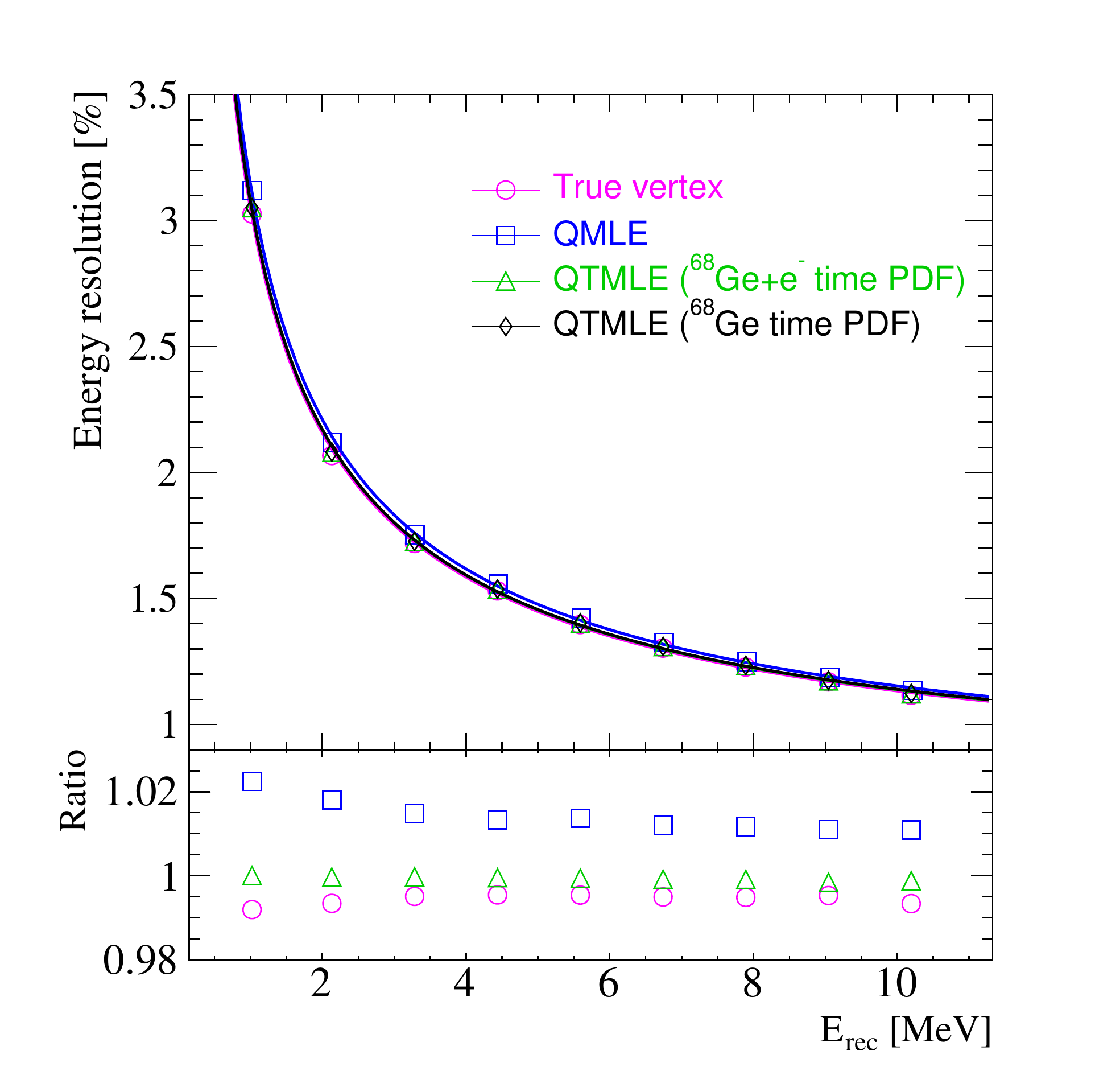}
    \caption{Comparison of the energy resolution with different vertex precision. 
    The bottom panel shows the ratio to the QTMLE method with $^{68}$Ge time PDF.
    The blue dots correspond to the energy resolution of QMLE, which has the worst vertex resolution. While the pink dots represent the energy resolution of QTMLE using true vertex, which is equivalent to an ideal vertex resolution of 0~mm. The energy resolution of QTMLE using $^{68}$Ge + electron time PDF is also shown and will be discussed in Sec.~\ref{sec:discussion}.}
\label{fig:ERecPrfm_Eres}
\end{figure*}

Given the likelihood functions from QMLE with charge information only and TMLE with time information only, it is straightforward to construct the charge and time combined maximum likelihood estimation (QTMLE) as Eqn.~\ref{eq:EqQTMLE}, by multiplying Eqn.~\ref{eq:EqQMLE} and Eqn.~\ref{eq:EqTMLE}.
\begin{equation}\small
\label{eq:EqQTMLE}
\begin{split}
    &\mathcal{L}(q_{1},q_{2},...,q_{N}; t_{1,r},t_{2,r},...,t_{N,r}|\mathbf{r},t_0,
    E_{\textrm{vis}}) = \\
      & \prod_{unfired}e^{-\mu_{j}} \prod_{fired} \left(\sum_{k=1}^{+\infty} P_{Q}(q_{i}|k)\times P(k, \mu_i)  \right) \\ 
      &\prod_{T-\textrm{valid}~hit} \left( \frac{\sum^{K}_{k=1} P_{T}(t_{i,r}|r, d_i, \mu^{l}_i, \mu^{d}_i, k)\times P(k, \mu^{l}_i+\mu^{d}_i)} {\sum^{K}_{k=1} P(k, \mu^{l}_i+\mu^{d}_i)}\right).
\end{split}
\end{equation}
Since QTMLE uses both the charge and time information of PMTs to constrain the vertex, it yields the best vertex reconstruction performance among the three methods, 
which can be seen from the left plot in Fig.~\ref{fig:VRecPrfm_res}. Across the whole energy range of interest, the vertex resolution of TMLE and QTMLE is on average about 56\% and 60\% better with respect to QMLE, respectively. 
Meanwhile, a more accurate vertex also leads to more accurate energy given the strong correlation between them. This can be seen in Fig.~\ref{fig:ERecPrfm_Eres}, where the energy resolution of QTMLE is much better comparing to QMLE. The resolution of the x, y, z, r components of the vertex is similar, with a discrepancy less than 4\%, as shown in Fig.~\ref{fig:VRecPrfm_XYZres}. Meanwhile the energy resolution is not sensitive to the resolution of $\theta$ and $\phi$ components, given the approximate spherical symmetry of the CD. Thus only the radius resolution is presented throughout this paper.

\begin{figure}[!ht]
\centering
    \includegraphics[width=\figwb\textwidth]{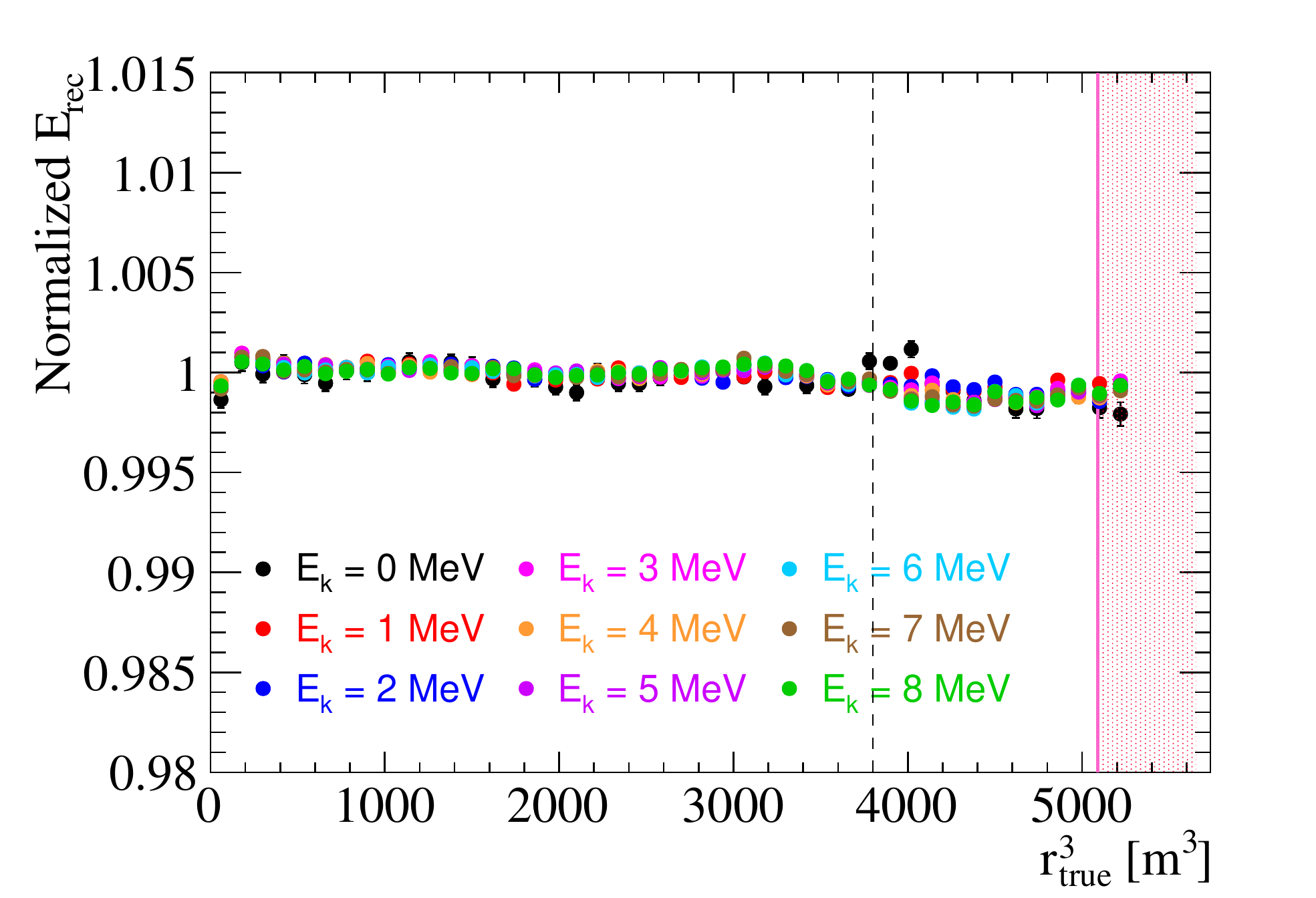}
\caption{Energy uniformity of QTMLE. The QTMLE method yields excellent energy uniformity. The residual energy non-uniformity is less than $0.23\%$ within the FV for all the different energies.}
\label{fig:ERecPrfm_Eunif}
\end{figure}

The uniformity of the reconstructed energy using QTMLE for positrons with different energies is shown in Fig.~\ref{fig:ERecPrfm_Eunif}. 
The QTMLE method yields excellent energy uniformity and the residual energy non-uniformity is less than 0.23\% inside the fiducial volume. 
Compared with the 0.17\% value in Ref. ~\cite{Huang_2021}, which uses true vertex and does not include any electronic effects, one can see that the impact of vertex inaccuracy and electronic effects on the energy non-uniformity is non-negligible but still under control.

\subsection{Discussion}
\label{sec:discussion}
As mentioned previously, the energy deposition process of positrons in LS usually consists of two parts: the kinetic energy and the annihilation with an electron emitting two gamma particles. 
During the construction of the expected nPE map of PMTs, Laser and $^{68}$Ge were used to mimic the two parts respectively. 
However, since the photons from Laser only contain the fast component, it can not be used to describe the photon timing profile of charged particles. 
Although electrons can mimic the kinetic energy part of positrons, mono-energetic electron sources are not available.
As a result, only the $^{68}$Ge source was used to construct the time PDF of PMTs in all the previous studies. Pseudo electron calibration data was produced to check the impact of the accuracy of the time PDF on the vertex and energy reconstruction. 
The QTMLE method was used and a new set of time PDF was constructed using weighted $^{68}$Ge + electron time PDF.
The reconstruction results were compared to those using $^{68}$Ge time PDF. 

The right plot in Fig.~\ref{fig:VRecPrfm_res} shows the comparison of the vertex resolution. 
The weighted $^{68}$Ge + electron time PDF is more accurate than the $^{68}$Ge time PDF, and the corresponding vertex resolution is about 8\% better.
Fig.~\ref{fig:ERecPrfm_Eres} shows the comparison of the energy resolution between the two cases. Despite the better vertex resolution from using $^{68}$Ge + electron time PDF, the energy resolution is almost the same as the one from using $^{68}$Ge time PDF.
 To check the impact of the accuracy of the vertex on the energy resolution, two additional cases, namely QMLE and QTMLE using true vertex, were also drawn in Fig.~\ref{fig:ERecPrfm_Eres}. 
The black dots correspond to the energy resolution of QMLE, which has the worst vertex resolution. 
While the pink dots represent the energy resolution of QTMLE using true vertex, which has an ideal vertex resolution of 0~mm. 
By comparing these cases, it is clear that better vertex resolution leads to better energy resolution. Meanwhile, comparing the default case of QTMLE using $^{68}$Ge time PDF to the ideal case of QTMLE using true vertex, the impact of vertex inaccuracy on the energy resolution is around 0.6\%.

\section{Conclusion}
\label{sec:sum}
High precision vertex and energy reconstruction are crucial for large liquid scintillator detectors such as JUNO, especially for the determination of the neutrino mass ordering. 
In this paper, a calibration data-driven simultaneous vertex and energy reconstruction method was proposed. 
The dependence of the refractive index on the photon propagation distance was calibrated to obtain more precise PMT time information.
More accurate and realistic time PDF of PMTs were constructed  to take into account the dependence on the vertex radius and photon propagation distance. The contribution to the time PDF from PMT dark noise was modeled in an analytical approach. 
With all these updates, a charge and time combined likelihood function was constructed to simultaneously reconstruct the vertex and energy of positrons. 
This method does not reply on MC simulation and obtains the expected PMT charge and time response directly from calibration data. 
It also naturally handles the strong correlation between vertex and energy. 
By combining the PMT charge and time information, the vertex resolution was improved by about 4\% (60\%) with respect to using the time (charge) information only. 
The vertex bias was reduced with the more accurate time PDF and less than 2~cm.
Better vertex resolution also leads to better energy resolution. The residual energy non-uniformity of this method is less than 0.5\% within the FV. Moreover, the impact of inaccurate vertex on the energy resolution is about 0.6\%.

\section*{Acknowledgements}
This work was partially supported by the National Key R\&D Program of China (2018YFA0404100), 
by the Strategic Priority Research Program of the Chinese Academy of Sciences (XDA10010100), 
by the National Natural Science Foundation of China (Grant No.12175257), 
and by the Science Foundation of High-Level Talents of Wuyi University (2021AL027).

\printbibheading[heading=bibintoc]
\printbibliography[heading=none]

\end{document}